\documentclass[11pt,a4paper]{article}

\usepackage{natbib}
\usepackage{epsfig}
\usepackage{setspace} 

\begin{document}

\title{Dynamics  and robustness of familiarity memory }

\author{J.M. Cortes$^1$, A. Greve, A.B. Barrett and M.C.W. van Rossum \\
Institute for Adaptive and Neural Computation\\
 School of Informatics, University of Edinburgh\\
 5 Forrest Hill, Edinburgh EH1 2QL, 
 UK \\
1. Correspondence to: Jesus M. Cortes. email: jcortes1@inf.ed.ac.uk}

\date{}
\maketitle

\begin{abstract} 

When one is presented with an item or a face, one can sometimes have a sense of
recognition without being able to recall where or when one has encountered it
before. This sense of recognition is known as familiarity. Following previous
computational models of familiarity memory we investigate the dynamical
properties of familiarity discrimination, and contrast two different familiarity
discriminators: one based on the energy of the neural network, and the other
based on the time derivative of the energy. We show how the familiarity signal
decays  after a stimulus is presented, and examine the robustness of the
familiarity discriminator in the presence of random fluctuations in neural
activity. For both discriminators we establish, via a combined method of
signal-to-noise ratio and mean field analysis, how the maximum number of
successfully discriminated stimuli depends on the noise level.


Abbreviations: SNR, Signal-to-Noise Ratio; FamE, Familiarity discrimination
based on Energy; FamS, Familiarity discrimination based on Slope. 
\end{abstract}

\section*{Introduction}

Recognition memory is supported by at least two different types of
retrieval processes: recollection and familiarity. While recollection
requires detailed information about an experienced event, familiarity
just distinguishes whether or not the stimulus was previously encountered.
A well known example is the encounter with a colleague during a conference:
one might recognize the person, but fail to remember the time and place
of an earlier meeting.

Familiarity memory is thought to have a very large capacity. In the
early 1970s, Standing and collaborators \citep{standing73} tested
the capacity in humans by presenting participants with a large number
(10,000) of images. Surprisingly, after this one-shot learning, participants
were able to successfully recognize most of the previously seen pictures,
suggesting that the capacity of recognition memory for pictures is
very large indeed.

Experimental psychologists have formulated \textit{dual-process} theories
which characterize the precise contribution of familiarity and recollection to
 recognition memory, for a review see \citep{yonelinasREV}.
Anatomically, researchers have proposed that  different brain areas  are engaged during
recollection and familiarity processing. Single item familiarity is believed
to be processed in the perirhinal cortex, whereas recollection is
believed to engage  the hippocampus, for a review see \citep{brown2001}.
Furthermore, electro-physiological studies using single cell recordings
in monkeys and rats \citep{brown87,brown98} report that about $30$
percent of neurons in the perirhinal cortex show increased activity
after presenting new compared to old stimuli. These neurons have been
interpreted as novelty detectors. However, this association between
the memory processes and brain area is still unclear and seems to
depend on the nature of the stimulus \citep{brown2005,yonelinas2003}.

Recent neuroimaging studies using, for example, event-related potentials
(ERPs) \citep{yonelinas2003}, have revealed that familiarity and recollection
have distinct temporal characteristics. Familiarity is linked to a frontal ERP modulation that occurs around $300$-$500$ms post-stimulus presentation, whilst   recollection is thought to evoke a parietal ERP modulation around $500$-$800$ms after stimulus presentation \citep{rugg1998,Greve2007}.
 Therefore, the
speed of processing of familiarity discrimination is more rapid than
recollection. Behavioral experiments provide further evidence for
this: if only very limited time is available for a recognition decision,
participants rely primarily on familiarity as opposed to recollection
processes  \citep{dosher1984}.

In the field of computational neuroscience, modeling of recollection
via attractor neural networks has a long history using auto-associator
Hopfield networks \citep{hopfieldPNAS,amitBOOK}. Familiarity discrimination has
only been studied much more recently \citep{bogaczHIPPO}. Computational
models of familiarity discrimination have a much higher storage capacity
for recognition than associative memory networks that perform associative
recall. For a wide range of conditions, Bogacz et al. showed that
the maximum storage is proportional to the number of synapses within
the network \citep{bogaczHIPPO}. This is much larger than the capacity
to recall, which is proportional to the square root of the number
of synapses (i.e.~the number of neurons in a fully connected network)
\citep{amitBOOK}. Intuitively this is easily understood;
familiarity needs to store just a single bit (familiar versus non-familiar)
per pattern, whereas to recall an event requires the retrieval of
the whole pattern (pattern completion).

In this paper we study how the dynamics of the network affects familiarity
discrimination. We compare two different familiarity discriminators:
familiarity based on Energy, FamE, which was previously introduced
by Bogacz et al \citep{bogaczHIPPO}, and a familiarity discriminator
which is the time derivative of FamE \citep{hopfieldPNAS}. From here
on, we will call this latter discriminator the slope, and label it
FamS.

We show in our model how the signal for both familiarity discriminators
decays very quickly after stimulus presentation, in concordance with
the results from neuroimaging \citep{rugg1998,Greve2007}. In addition,
to investigate the robustness of familiarity detection, we study how
it is affected by random fluctuations that are ubiquitously present
in the nervous system. As in models of attractor neural networks \citep{amitBOOK},
this external source of noise is taken to be independent of the learned
patterns and is controlled by a \textit{temperature} parameter. 

\section*{Two familiarity discriminators}

We consider a network of $N$ binary neurons, each with an activity
$s_{i}(t)=\pm1$, the two states corresponding respectively to firing
and not firing. The complete network activity is characterized by
$\mathbf{s}(t)$. Any two neurons are connected by synaptic weights
$w_{ij}$. As standard in artificial network models, the network has
a learning phase in which it encodes $M$ stimuli $\mathbf{x^{\rho}}\equiv\{ x_{i}^{\rho}\}_{i=1}^{N}$
($\rho=1\ldots M$) in its weights using a Hebbian learning rule \begin{equation}
w_{ij}=\frac{1}{N}\sum_{\rho=1}^{M}x_{i}^{\rho}x_{j}^{\rho}.\label{ws}\end{equation}
 It can be shown that this learning rule is optimal in the limit of
large $N,M$ (unpublished results). During the subsequent test phase,
the network's performance is evaluated. At $t=0$, the probe stimulus
$\hat{\rho}$ (which is either a familiar or novel stimulus) is loaded
into the network, $\mathbf{s}(t=0)={\bf x^{\hat{\rho}}}$. 

To define the network dynamics we assume that each neuron is updated precisely
once, probabilistically and asynchronously, in each unit of time. 
(The exact duration that a time unit in the model corresponds to is hard to 
extract by comparing the model to, say, ERP data given the additional delays 
present in the biology, but it should probably be on the order of 10..100ms.)
As standard in artificial neural networks, and in
analogy with magnetic systems in physics, the random fluctuations
are controlled by a temperature parameter $T$. These so-called Glauber
dynamics have been extensively studied in many different stochastic
systems, for instance \citep{marroBOOK}. The probability distribution, after update, is given 
then by
\begin{eqnarray}
P\{s_i(t+1) = \pm 1 \} =  \frac{1} {1+\exp[\mp 2 \beta h_i(t)]},\label{tr}
\end{eqnarray}
where $\beta\equiv1/T$ is the inverse temperature parameter, and $h_{i}(t)\equiv\sum_{j=1}^{N}w_{ij}s_{j}(t)$ is the total
presynaptic current to the neuron $i$. Accordingly, for low temperature, the
noise is small and there is a strong positive correlation between
the input current $h_i$ and the output $s_{i}$, whilst for high
temperature the output of a node is dominated by noise and is more
or less independent of its input.

The energy in the network at time $t$ is defined as \begin{equation}
E(t)\equiv-\sum_{ij}w_{ij}s_{i}(t)s_{j}(t).\label{E}\end{equation}
As was previously reported in \citep{bogaczHIPPO}, the energy $E(t=0)$
is able to effectively discriminate between old and novel stimuli. 
As we explain later, this energy is of order $-(N+M)$ for learned stimuli, but of order $-M$
for novel stimuli. Consequently, the energy or familiarity for old
and novel stimuli are macroscopically different (they differ by order
$N$, while the std.dev.$=\sqrt{2M}$) and the difference can thus
be used as a familiarity discriminator. We call this discriminator
FamE.

However, the use of the energy is only one possible approach to model
familiarity discrimination. The time derivative, or slope, of the energy $S=\frac{dE(t)}{dt}$
can also be used as a familiarity discriminator. It indicates how
quickly the network's energy changes, when either a novel or old stimulus
is presented. Interestingly, this familiarity measure was originally
proposed by Hopfield in his seminal 1982 paper \citep{hopfieldPNAS},
but to the best of our knowledge it has never received further exploration.
We call this discriminator FamS.

 For convenience, we shall express the energy and the slope as functions
of the $M$-dimensional vector $\mathbf{m}(t)\equiv\left\{ m^{\rho}(t)\right\} _{\rho=1}^{M}$,
the overlaps between the current network activity and each of the
stored patterns. The components of this overlap vector are defined
by \begin{equation}
m^{\rho}(t)\equiv\frac{1}{N}\sum_{i=1}^{N}x_{i}^{\rho}s_{i}(t).\label{ms}\end{equation}
 Assuming the Hebbian learning rule (\ref{ws}), the energy (\ref{E})
in terms of the overlaps is given by \begin{equation}
E(t)=-N\sum_{\rho=1}^{M}\left[m^{\rho}(t)\right]^{2},\label{Em}\end{equation}
whilst the slope (first derivative of the energy) is given by \begin{equation}
S(t)=-2N\sum_{\rho=1}^{M}m^{\rho}(t)\frac{dm^{\rho}(t)}{dt},\label{S}\end{equation}
 and is thus proportional to the time derivative $dm^{\rho}(t)/dt$ of the overlaps.

\section*{Dynamics of familiarity discrimination}

To mathematically address the network dynamics we assume the mean field
approximation, i.e.~$s_{i}\approx  \langle s_i  \rangle$. Under this
approximation one obtains from equation (\ref{tr}), the dynamical equations
for the overlaps (\ref{ms}):
 \begin{equation}
\frac{dm^{\rho}(t)}{dt}=-m^{\rho}(t)+\frac{1}{N}\sum_{i=1}^{N}x_{i}^{\rho}\tanh[\beta\sum_{\nu=1}^{M}x_{i}^{\nu}m^{\nu}(t)].\label{Dms}\end{equation} The
mean field formulation provides an accurate description of the dynamics
of the system provided the temperature is not too high (see below).

\begin{figure}
\centerline{ \psfig{file=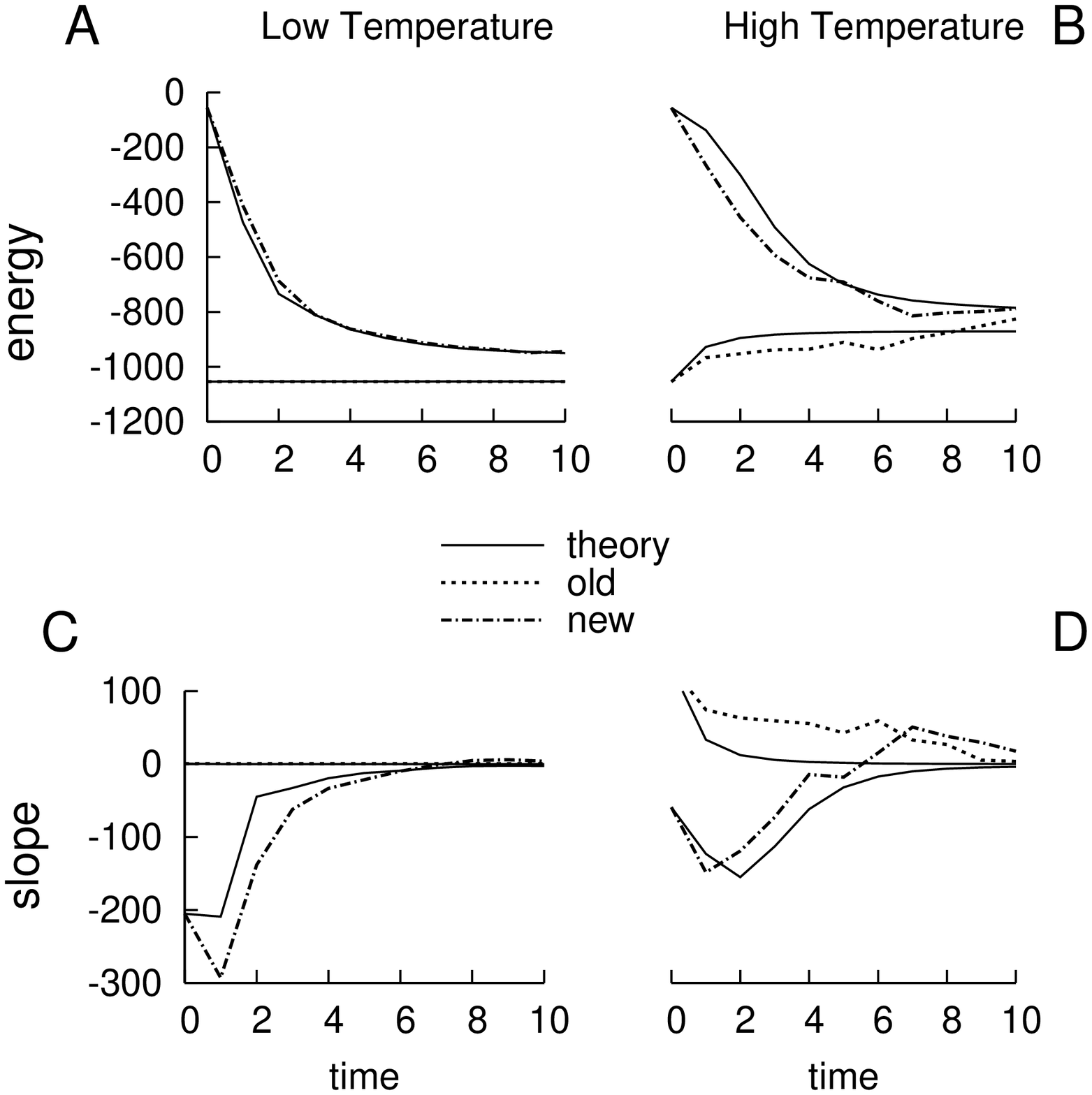,width=12cm} }

\caption{\textbf{Temporal persistence of discrimination by familiarity.} For
different values of the temperature parameter, $T=0.20$ on the left
and $T=0.60$ on the right, we simulate a network of $N=1000$ neurons
and $M=50$ uncorrelated patterns. Both FamE and FamS can discriminate
between novel and old stimuli during a short period post stimulus
presentation. After this, the slope begins to tend to zero, indicating
that the activity has converged to one of the stored stimuli. This
is due to the well-known pattern completion dynamics that occurs in
attractor neural networks. One unit of time is defined
as the time taken to update the whole population of neurons in the
network.}

\label{fig1} 
\end{figure}

Knowing the dynamics, we focus on the time evolution of the
two discriminators, energy and slope, defined in the previous section.
To measure the temporal persistence, Figure \ref{fig1} illustrates
the time evolution of FamE and FamS when tested with novel or old
stimuli. We compare the time evolution by simulations with Glauber dynamics
given by equation (\ref{tr}), and by using the mean field dynamical equations
(\ref{Dms}).

\vspace{0.5cm}
 \textit{(FIGURE 1 HERE)} \vspace{0.5cm}

Figure \ref{fig1} A and B, shows how the energy associated with old stimuli is
much lower than for new stimuli. However, after a short transient of $4$-$5$
units of time, both signals become similar to each other, i.e.~familiarity
discrimination based on energy deteriorates rapidly post stimulus presentation.

Like the energy, the slope also shows a transient signal when the network is
presented with a novel vs old stimulus, figure \ref{fig1} graphs C and D. For
low temperature, the slope for old stimuli is practically zero. This can be
easily interpreted. An old stimulus corresponds to one of the local minima
(attractors) of the energy landscape. Because the temperature is low, and
therefore the system is not receiving any external perturbation, the energy does
not change, and its time derivative is practically zero. Similar to the energy,
the slopes associated with old and new stimuli show significant differences 
immediately  after stimulus presentation, but this difference diminishes shortly
thereafter.

Summarizing, both discriminators can distinguish old from new stimuli
immediately after stimulus presentation, but after a very short transient
(of the order of five time units), the discrimnation ability disappears.
The slope tends to zero as time progresses because the network evolves
towards a fixed point and becomes stationary (i.e.~$S\approx0$). Though
measures to discriminate spurious from non-spurious attractor states
have been proposed \citep{robins2004}, such measures do not directly
translate into a discrimination between old and novel stimuli.

\section*{Robustness of the familiarity discriminators}

To examine the performance of the two familiarity discriminators introduced
in the previous section, we quantify the discriminability between the
network responses to either new or old stimuli by the signal-to-noise
ratio (SNR). Assuming two Gaussian probability distributions, $\mathcal{N}[\mu_{\mathrm{new}},\sigma_{\mathrm{new}}^{2}]$
and $\mathcal{N}[\mu_{\mathrm{old}},\sigma_{\mathrm{old}}^{2}]$,
associated with new and old stimuli, we define \begin{equation}
\mathrm{SNR}=\frac{\left|\mu_{\mathrm{new}}-\mu_{\mathrm{old}}\right|}{\sqrt{{\frac{1}{2}\sigma}_{\mathrm{new}}^{2}+{\frac{1}{2}\sigma}_{\mathrm{old}}^{2}}}.\label{snr}\end{equation}
 To check that the distributions are indeed Gaussian, we repeated
the simulation of Fig.~\ref{fig1} $100$ times and computed the
probability distributions. For both FamE and FamS the 4th moments
of their distribution satisfied $\langle x^{4}\rangle=\int P(x)x^{4}dx=\mu^{4}+6\mu^{2}\sigma^{2}+3\sigma^{2}\sigma^{2}$,
with a relative error smaller than $5\%$, (where $\mu=\langle x\rangle$
denotes the mean and $\sigma^{2}=\langle x^{2}\rangle-\langle x\rangle^{2}$
the variance), indicating that the distributions are well approximated
by Gaussians.

We address here how random fluctuations in neural activity (independent
of the learned patterns) affect the performance of the familiarity
discriminators. We study the effect of temperature at two different
time points, $t=0$ and $t=1$. As stated above, time is defined such
that in one unit, all neurons are asynchronously updated once. The
choice of $t=1$ is not special; we just study the network properties
at this time to gain understanding as to how the network evolves.

The results are illustrated in figure \ref{fig2}. Immediately after stimulus
($t=0$), we observe that FamE is independent of the temperature value (figure
\ref{fig2}.A), whilst FamS has a non-linear dependence on the temperature
(figure \ref{fig2}.C). For high temperature, FamS performs better as a
familiarity discriminator. This finding can be intuitively understood. The
energy and its time derivative can be separated into signal and noise
contributions. The signal for the slope is proportional to the rate of change of
the energy, and therefore proportional to the rate of change of the overlap
between the network activity and the stimulus. At low temperatures, the signal
associated with an old stimulus is very low as the overlap with the stimulus is
almost invariant. Contrarily, at higher temperature, the overlap with old
stimuli changes very quickly; it decays from $1$ to $0$, and consequently the
slope-signal relationship increases considerably (the higher temperature, the
higher signal for FamS). The noise component for the slope, although dependent
on $T$, is similar for both old and novel stimuli. As a result the main
temperature dependence stems from the signal term.  In figure \ref{fig2}, the
case of $T>1$ is not explicitely studied because in this region  the network can
not retrieved any of the learned patterns, i.e. the only stable solution is  
$\mathbf{m}=0$, what is so-called   \textit{paramagnetic} or \textit{non-memory}
solution  \citep{amitBOOK}.

\vspace{0.5cm}
 \textit{(FIGURE 2 HERE)} \vspace{0.5cm}

In contrast to time $t=0$, at time $t=1$, post stimulus presentation,
both discriminators FamE and FamS show a similar breakdown in discrimination
for increased temperature (figure \ref{fig2}.F). In the next section,
we analytically study the maximum storage capacity for both FamE and
FamS at time $t=0$. The results are in agreement with the simulations.
For $t=1$, the mean field predictions, however, do not reproduce
the network simulations. To study such situations (which we do not
explicitly deal with here), one would need to use other techniques,
for example, generating functional analysis \citep{coolenDYN}.

\section*{Maximum storage capacity}

When the number of stimuli encoded in the weights increases, the SNR
decreases. One can define the storage capacity (or maximum number
of stimuli encoded in the learning rule and successfully discriminated)
as the point where the SNR drops below one. This gives the maximum
number of stimuli $M_{\mathrm{max}}$ that can be encoded in the network.
In this section we present explicit calculations for both discriminators
FamE and FamS for time $t=0$.

\subsection*{Storage capacity of FamE}

Let $\rho=\hat{\rho}$ label an old stimulus presented to the network.
As is common in these calculations \citep{herzBOOK}, we separate the
sum appearing in equation (\ref{E}) into a signal ($\rho=\hat{\rho}$)
plus noise contributions. The latter is determined by interference
from previously stored stimuli ($\rho\neq\hat{\rho}$). From equation
(\ref{Em}) it follows that the energy associated with old stimuli is distributed
as \begin{equation}
E_{\mathrm{old}}\left(t\right)=-N[m^{\hat{\rho}}\left(t\right)]^{2}-N\sum_{\rho\neq\hat{\rho}}[m^{\rho}(t)]^{2}.\end{equation}
 The first term on the right hand side is the signal and the second
one the noise contribution. At $t=0$, we obtain $m^{\hat{\rho}}(t=0)=1$
because the pattern $\hat{\rho}$ was an old stimulus. As for large
$N$ the central limit theorem applies, the overlaps with the other
patterns $\rho\neq\hat{\rho}$ have a Gaussian distribution with $0$
average and variance $1/N$ \citep{amitAP}. Accordingly, we can easily
compute the expected value and the variance for the energy. Using
that the sum of two Gaussian distributed variables is again a Gaussian
distribution, \begin{equation}
E_{\mathrm{old}}(t=0)\in\mathcal{N}[-(N+M),2M].\label{Eoldt0}\end{equation}
Analogously, the energy
for novel stimuli is distributed as \begin{equation}
E_{\mathrm{new}}(t=0)\in\mathcal{N}[-M,2M].\label{Enewt0}\end{equation}
From equation (\ref{snr}) we obtain $\mathrm{SNR}=\sqrt{N^{2}/(2M)}$,
in agreement with the simulations (see figure \ref{fig2}.E). Equivalently,
the maximum storage capacity, (the $M$ for which $\mathrm{SNR}=1$), is given by \begin{equation}
M_{\mathrm{max}}[\mathrm{FamE},t=0]=\frac{N^{2}}{2},\label{MmaxEt0}\end{equation}
and thus the storage is of order $N^{2}$, which has been reported in previous
computational models using FamE \citep{bogaczHIPPO}.

\subsection*{Storage capacity of FamS}

Following the same strategy applied to FamE to FamS, we are able to
separate signal ($\rho=\hat{\rho}$) and noise ($\rho\neq\hat{\rho}$)
terms for the slope. At the instant of the stimulus presentation ($t=0$),
we substitute equation (\ref{Dms}) in equation (\ref{S}). Next,
we apply the central limit theorem, which is a good approximation
for large $N$. It ensures that the sum over the different sites $i$
of the noise contribution $\sum_{\rho\neq\hat{\rho}}x_{i}^{\rho}m^{\rho}$
appearing inside the $\tanh$ function, is equivalent to the average over
a Gaussian noise with mean $0$ and variance $\alpha\equiv M/N$,
the network \textit{load}. Using these considerations, it is straightforward
to obtain  \begin{equation}
S_{\mathrm{old}}(t=0)\in\mathcal{N}\left[2N\left(1-I_{1}-I_{2}\right)+2M,8M\right],\label{Soldt0}\end{equation}
 and for novel stimuli \begin{equation}
S_{\mathrm{new}}(t=0)\in\mathcal{N}\left[-2NI_{3}+2M,8M\right].\label{Snewt0}\end{equation}
 The integrals $I_{1}$, $I_{2}$ and $I_{3}$ appearing in equations
(\ref{Soldt0}) and (\ref{Snewt0}) are \begin{eqnarray}
I_{1}(\alpha,\beta) & \equiv & \int\frac{\mathrm{d}z}{\sqrt{2\pi}}\exp\left(-z^{2}/2\right)\tanh\left(\beta+\beta\sqrt{\alpha}z\right),\nonumber \\
I_{2}(\alpha,\beta) & \equiv & \int\frac{\mathrm{d}z}{\sqrt{2\pi}}\exp\left(-z^{2}/2\right)\tanh\left(\beta+\beta\sqrt{\alpha}z\right)\sqrt{\alpha}z,\label{I1I2I3}\\
I_{3}(\alpha,\beta) & \equiv & \int\frac{\mathrm{d}z}{\sqrt{2\pi}}\exp\left(-z^{2}/2\right)\tanh\left(\beta\sqrt{\alpha}z\right)\sqrt{\alpha}z,\nonumber \end{eqnarray}
 where $\beta\equiv1/T$ is the inverse temperature. From equations
(\ref{Soldt0}) and (\ref{Snewt0}) it follows that $\mathrm{SNR}=\sqrt{N^{2}/(2M)}[1-I_{1}(\alpha,\beta)-I_{2}(\alpha,\beta)+I_{3}(\alpha,\beta)]$,
which can be computed numerically. The results are represented in
figure \ref{fig2}.E. The expected values used in the signal-to-noise
ratio calculation (figure \ref{fig2}.C) fits well with the simulations.
However, the theoretical predictions for the variance of both FamS(old)
and FamS(new) equals $8M$, independent of temperature, which is in
disagreement with the simulations. Therefore, the signal-to-noise
ratio calculation disagrees with simulations for high temperatures
(figure \ref{fig2}.E). See the appendix for a more detailed calculation
of how the mean field prediction is affected by high temperatures.

\vspace{0.5cm}
 \textit{(FIGURE 3 HERE)} \vspace{0.5cm}

The maximum storage for FamS is again obtained by solving SNR$=1$,
which yields \begin{equation}
M_{\mathrm{max}}[\mathrm{FamS},t=0]=\frac{N^{2}}{2}(1-I_{1}(\alpha_{\mathrm{max}},\beta)-I_{2}(\alpha_{\mathrm{max}},\beta)+I_{3}(\alpha_{\mathrm{max}},\beta))^{2}.\label{MmaxSt0}\end{equation}
 Because the integrals $I_{1}(\alpha,\beta)$, $I_{2}(\alpha,\beta)$ and $I_{3}(\alpha,\beta)$ depend on $M$,
this expression does not give us $M_{\mathrm{max}}$ explicitly. The
dependence on $N$ is more complicated than for other computational
models of familiarity discrimination \citep{bogaczHIPPO}, (and in
particular for FamE above), for which the maximum storage capacity
is directly proportional to $N^{2}$. Interestingly, $M_{\mathrm{max}}$
for FamS at $t=0$ is dependent on the temperature, whilst FamE is
completely independent of temperature (recall figure \ref{fig2},
graphs A and C).

In the two limits $T=0$ and $T\rightarrow\infty$ we can perform
the integrals in equation (\ref{MmaxSt0}) to obtain $M_{\mathrm{max}}$
explicitly. For $T=0$, the integrals (\ref{I1I2I3}) can be computed
using \begin{eqnarray}
\lim_{\beta\to\infty}\int\frac{\mathrm{d}z}{\sqrt{2\pi}}\exp\left(-z^{2}/2\right)\tanh\left(\beta\left[az+b\right]\right) & = & \mathrm{erf}\left(\frac{b}{\sqrt{2}a}\right),\nonumber \\
\lim_{\beta\to\infty}\int\frac{\mathrm{d}z}{\sqrt{2\pi}}\exp\left(-z^{2}/2\right)\tanh\left(\beta\left[az+b\right]\right)z & = & \sqrt{\frac{2}{\pi}}\exp\left(-\frac{b^{2}}{2a^{2}}\right),\label{appT0}\end{eqnarray}
 giving $\lim_{\beta\to\infty}I_{1}(\alpha,\beta)=\mathrm{erf}\left(\frac{1}{\sqrt{2\alpha}}\right)$,
$\lim_{\beta\to\infty}I_{2}(\alpha,\beta)=\sqrt{\frac{2\alpha}{\pi}}\exp\left(-\frac{1}{2\alpha}\right)$
and $\lim_{\beta\to\infty}I_{3}(\alpha,\beta)=\sqrt{\frac{2\alpha}{\pi}}$. Here,  $\mathrm{erf}(x)$ is the error function
$\mathrm{erf}\left(x\right)\equiv\frac{2}{\sqrt{\pi}}\int_{0}^{x}\exp\left(-u^{2}\right)\mathrm{d}u$.
Therefore at $T=0$ equation (\ref{MmaxSt0}) becomes \begin{equation}
M_{\mathrm{max}}=\frac{N^{2}}{2}\left(1-\mathrm{erf}\left(\sqrt{\frac{N}{2M_{\mathrm{max}}}}\right)+\sqrt{\frac{2M_{\mathrm{max}}}{\pi N}}\left[1-\exp\left(-\frac{N}{2M_{\mathrm{max}}}\right)\right]\right)^{2}.\label{MST0}\end{equation}
 In figure \ref{fig3} we plot, as a function of $N$, the ratio of
the initial zero temperature storage for FamS and FamE. We see that
although FamS performs slightly worse than FamE, both storage capacities
grow proportional to $N^{2}$. By way of example, for $N=1000$, we
see $M_{\mathrm{max}}[\mathrm{FamS},t=0,T=0]\approx96\% M_{\mathrm{max}}[\mathrm{FamE},t=0]$,
i.e.~the capacities are almost identical. 

In the other limit that $T\rightarrow\infty$, random fluctuations
in neural activity dominate the network dynamics. All the integrals
of (\ref{I1I2I3}) are zero, and hence $M_{\mathrm{max}}[\mathrm{FamS},t=0]\approx M_{\mathrm{max}}[\mathrm{FamE},t=0]$.
That is, in this limit, the theoretical maximum storage is the same
for both FamS and FamE, and is independent of $T$.

\section*{Discussion}

Familiarity describes a retrieval process that supports recognition memory by
providing a feeling that something has been encountered before. Numerous
empirical studies have investigated familiarity processes in humans
\citep{yonelinasREV} and non-humans \citep{brown98}. Recently, some neuronal
networks modeling familiarity discrimination have also been proposed
\citep{bogaczHIPPO}. However, no computational work has addressed the dynamics of
familiarity discrimination, which is relevant when comparing these models to
experiments. Furthermore, we have studied how  noise 
affects the familiarity performance.

We have compared the energy discriminator (FamE) used by \citep{bogaczHIPPO}
to its time derivative, the slope (FamS). Interestingly,
the FamS discriminator was already suggested by Hopfield in his seminal
work \citep{hopfieldPNAS}. An interesting consequence is that the
original Hopfield model can be used to model both recollection (stationary
properties of the retrieval dynamics) and familiarity (transient dynamics
after the stimulus presentation). The slope discriminator (FamS) is
affected by the temporal dependency of the energy discriminator (FamE).
In other words, the slope discriminator captures the fact that the
speed of discrimination is predictive for the discrimination outcome
per se.

For both discriminators the familiarity signals decay quickly after 
stimulus presentation and are detectable only for a short period of time. This
can be compared to the speed of recollection. Assuming that recollection
memories correspond to attractors in the Hopfield model, recollection
information only becomes available once the attractor state is reached. By that
time, the slope is zero, and the energy difference is very small. Thus the
experimentally  observed timing difference of familiarity and recollection
follows naturally from our model.

The storage capacity of these familiarity discriminators
is much larger (proportional to $N^{2}$) compared to recollection (proportional
to $N$), we demonstrated that this capacity
is dependent on the \textit{temperature}. We have presented a detailed
derivation of the maximum storage immediately after stimulus presentation
($t=0$). We have shown that for low temperature, the storage capacity
related to FamS is lower than that for FamE, but still scales with the
number of synapses, e.g.~for $N=1000$, the slope gives a storage capacity $96\%$ as good as the energy.  
For high temperatures the difference between the storage capacities of 
FamS and FamE is negligible (the storage capacity for both is
the approximately, $N^{2}/2$). 

Interestingly, this means that the performance of FamS improves as one goes to 
the high temperature regime, a fact which is a priori counterintuitive,
especially given how the temperature affects \textit{recollection} in
Hopfield nets \citep{amitBOOK}, i.e.~the higher the temperature, the
worse the recollection performance. 
However, after some time steps, our simulations
(figure \ref{fig2}.F) show that, for both FamE and
FamS, high noise levels produce a stochastic disruption of the discrimination,
decreasing the SNR and the performance of familiarity, a concurrence
with the dynamics of recollection.

\subsection*{Acknowledgments}

The authors acknowledge Rafal Bogacz (Univ.~Bristol) and David Donaldson
(Univ.~Stirling) for helpful discussions and financial support from
EPSRC (project Ref.~EP/CO 10841/1), HFSP (project Ref.~RGP0041/2006)
and from the Doctoral Training Center in Neuroinformatics at the University
of Edinburgh.

\subsection*{Appendix: Mean field validity dependence on temperature}

\begin{figure}
\centerline{ \psfig{file=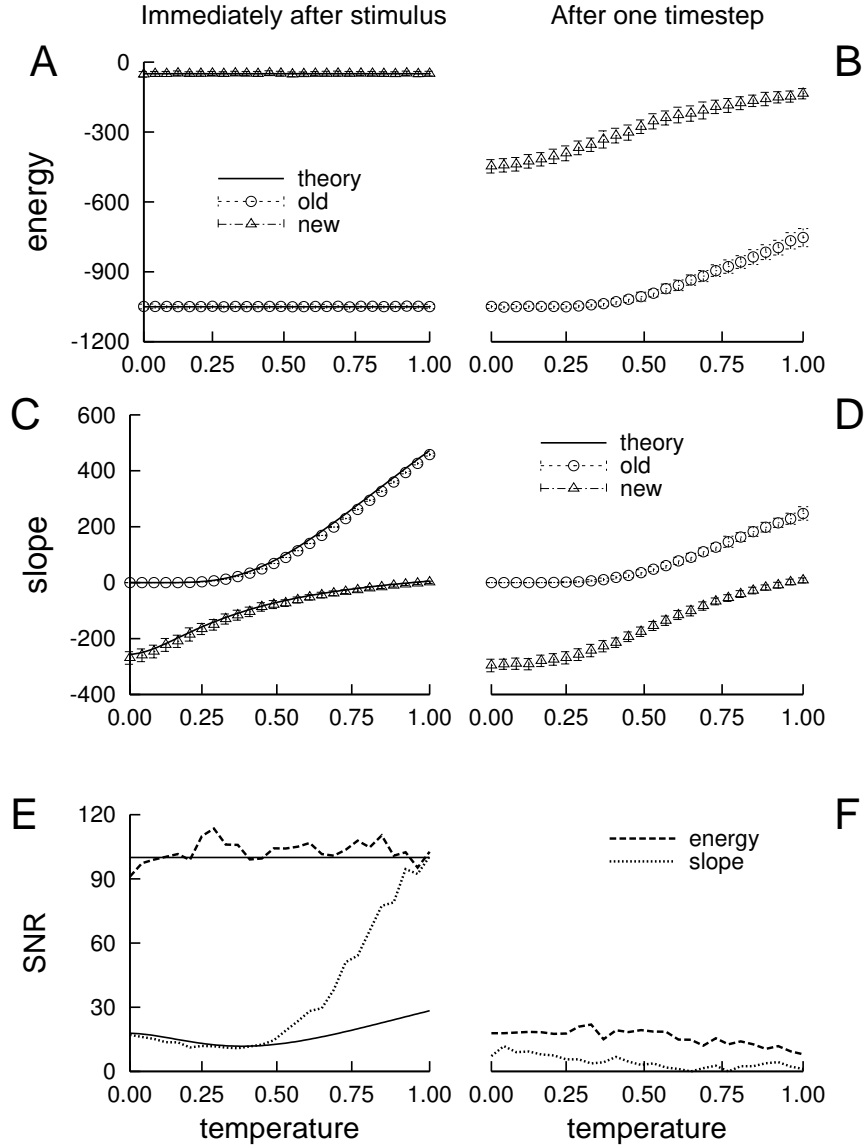,width=12cm} }

\caption{ \textbf{Robustness of discrimination by familiarity.} Immediately
after stimulus presentation, left graphs, FamE is independent of temperature,
whereas FamS is enhanced if the temperature parameter increases. After
one timestep, right graphs, both FamE and FamS deteriorate for high
values of temperature. We represent the values of the energy (top
graph) and the slope (middle) with the standard deviation. On the bottom,
each point in the curves corresponds with a fixed value of temperature,
in which we compute the SNR concerning the probability distributions
of the network responses towards both familiar and novel stimuli.
These simulations correspond with averaging over $100$ runs of a
network with $N=1000$ neurons and $M=50$ uncorrelated patterns.
Black solid lines are the theoretical predictions (see text for details). }

\label{fig2} 
\end{figure}

\begin{figure}
\centerline{ \psfig{file=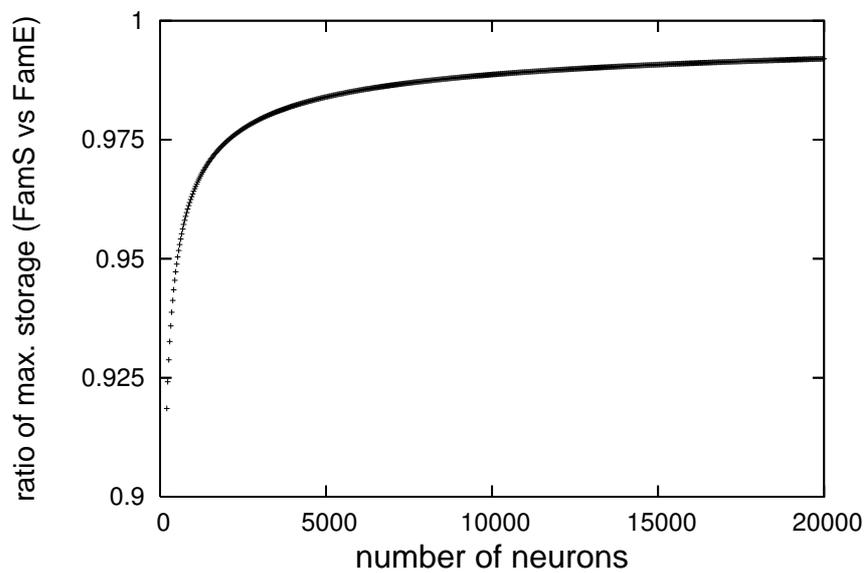,width=12cm} }

\caption{\textbf{Ratio of initial storage capacities at zero temperature.}
The storage of discriminator FamS is obtained by numerical solution
of equation (\ref{MST0}) as a function of the number of neurons $N$.
This is normalized by the storage for FamE (\ref{MmaxEt0}), to obtain
a ratio of the performances of the two discriminators.}

\label{fig3} 
\end{figure}

To compute the slope in equation (\ref{S}), we need an analytic expression
for $dm^{\rho}/dt$, or equivalently, given the
definition (\ref{ms}), we have to compute the derivative $ds_{i}/dt$,
which is governed by the Glauber dynamics given by (\ref{tr}), see
\citep{marroBOOK} for more detailed situations. Given $s_{i}(t)$,
the Glauber dynamics give an uncertainty in $s_{i}(t+1)$, such that
\begin{equation}
\mathrm{Var}[s_{i}(t+1)|\{ s_{j}(t)\}]=\mathrm{sech}^{2}(\beta h_{i}(t))\,,\end{equation}
 which implies \begin{equation}
\frac{ds_{i}}{dt}=\tanh(\beta h_{i})-s_{i}+\mathcal{O}(\mathrm{sech}(\beta h_{i}))\,.\label{err}\end{equation}
 We use this result to find the error induced in our calculation of
$S_{\mathrm{new}}$. When a new pattern is presented, the
$m^{\rho}$ are all of order $N^{-1/2}$. This implies that the local
fields, defined as $h_{i}\equiv\sum_{\rho}x_{i}^{\rho}m^{\rho}$,
are of order $\sqrt{\alpha}\equiv\sqrt{M/N}$. Hence, by equations
(\ref{ms}) and (\ref{err}), the error in our calculation of $dm^{\rho}/dt$
is given by \begin{equation}
\mathrm{Error}\left(\frac{dm^{\rho}}{dt}\right)=\mathcal{O}\left(\frac{1}{\sqrt{N}}\mathrm{sech}\left(\beta\sqrt{\alpha}\,\right)\right)\,,\end{equation}
 for each $\rho$. Finally, by (\ref{S}), we conclude that \begin{equation}
\mathrm{Error}\left(S_{\mathrm{new}}\right)=\mathcal{O}\left(\sqrt{M}\,\mathrm{sech}\left(\frac{1}{T}\sqrt{\frac{M}{N}}\,\right)\right)\,.\end{equation}
 Since $\mathrm{sech}(x)$ decays exponentially with large $x$, but
is of order 1 for small $x$, the error in our calculation of $S_{\mathrm{new}}$,
coming from the mean field approximation, is only going to be negligible
in the limit in which $(1/T)\sqrt{M/N}$ is large. This explains why
there is a growing discrepancy between theory and simulation as the
temperature $T$ is increased (see figure \ref{fig2}.E).

\newpage{}


\newpage{} %

\newpage{} %

\newpage{} %

\end{document}